\documentclass[aps,prl,groupedaddress,twocolumn]{revtex4-1}
\usepackage{graphicx,amsmath,amsfonts,xcolor,bm,braket,mathptmx}

\begin{document}

\title{Achieving the ultimate quantum timing resolution}

\author{V. Ansari$^{1,2}$}
\email{vansari@stanford.edu}
\author{B. Brecht$^1$}
\author{J. Gil-L\'opez$^1$}
\author{J. M. Donohue$^{3}$}
\author{J.~\v{R}eh\'a\v{c}ek$^{4}$}
\author{Z.~Hradil$^{4}$}
\author{L.~L. S\'{a}nchez-Soto$^{5,6}$}
\author{C. Silberhorn$^{1}$}

\affiliation{$^1$Integrated Quantum Optics, Paderborn University, 33098 Paderborn, Germany}
\affiliation{$^2$E. L. Ginzton Laboratory, Stanford University, 348 Via Pueblo Mall, Stanford, California 94305, USA}
\affiliation{$^3$Institute for Quantum Computing, University of Waterloo, Waterloo, Ontario N2L 3G1, Canada}
\affiliation{$^4$Department of Optics, Palack\'{y} University, 771 46 Olomouc, Czech Republic}
\affiliation{$^5$Departamento de \'{O}ptica, Facultad de F\'{\i}sica, Universidad Complutense, 28040 Madrid, Spain}
\affiliation{$^6$Max-Planck-Institute f\"{u}r die Physik des Lichts, 91058 Erlangen, Germany}

\date{\today}

\begin{abstract}
Accurate time-delay measurement is at the core of many modern technologies. Here, we present a temporal-mode demultiplexing scheme that achieves the ultimate quantum precision for the simultaneous estimation of the temporal centroid, the time offset, and the relative intensities of an incoherent mixture of ultrashort pulses at the single-photon level. We experimentally resolve temporal separations ten times smaller than the pulse duration, as well as imbalanced intensities differing by a factor of $10^{2}$. This represents an improvement of more than an order of magnitude over the best standard methods based on intensity detection. 
\end{abstract}


\maketitle

\section{Introduction}

The measurement of the time delay between two clocks is of paramount importance for many applications, from  navigation and global positioning~\cite{Droste:2015aa} to tests of general relativity~\cite{Chou:2010aa}, long baseline interferometry~\cite{Krehlik:2017aa}, optical coherence tomography~\cite{Fujimoto:1995aa}, and gravitational wave detection~\cite{Graham:2013aa}, to cite but a few. With optical pulses emitted from coherent or partially coherent sources, the timing information can be measured through established interferometric methods and conventional photo-detection, such as Fourier-transform interferometry~\cite{Weiner:2009aa}. Distance information can be extracted from timing information using the time-of-flight principle~\cite{Hansard:2013aa}, which detects reflections off of distant objects. In these cases and others, the main goal of a timing measurement is to estimate specific properties of a received signal consisting of multiple pulses, such as relative time delays, centroids, and relative intensities, and not necessarily full temporal profile reconstruction.

In many settings, the optical pulses being measured share little or no coherence. This happens with, for example, remote clocks (e.g GPS), incoherent excitations in biological samples, condensed matter physics, and astronomical observations~\cite{Maddaloni:2013aa}. In the absence of coherence, interferometric methods like FROG~\cite{Trebino:2002aa} and SPIDER~\cite{walmsley2009characterization} cannot be exploited, and the estimation precision of tools that directly measure temporal intensity, such as streak cameras~\cite{Chang:2011aa}, and time-to-space conversion~\cite{Nuss:1994aa}, is reduced dramatically (see Appendix A).

In the spatial domain, this problem has been dubbed as Rayleigh's curse~\cite{Tsang:2016aa}. In our context it can be formulated as the limits in  estimating the temporal separation $\tau$ between optical pulses~\cite{Donohue:2018aa}: For intensity-only direct-detection schemes in the instructive case of two mutually incoherent pulses with equal intensities, the information gained per photon detected (quantified by the Fisher information) decreases quadratically with $\tau$. This implies that the variance of the estimation of $\tau$ diverges as the pulse separation approaches zero, as can be formalized through the Cram\'er-Rao lower bound (CRLB)~\cite{Kay:1993aa}.

Employing appropriate strategies (such as homodyne detection~\cite{Lamine:2008aa}), the measurable timing sensitivity can be enhanced by a factor $1/\sqrt{N}$, where $N$ is the mean total photon number of photons measured in the experiment during the detection time. This is the famous standard quantum limit~\cite{Caves:1980aa}, which can be even surpassed to the ultimate Heisenberg scaling $1/N$~\cite{Giovannetti:2004aa}. Detecting more photons, however, is not always possible; in many photon-starved applications, such as astronomy or biological imaging, where longer measurement times suffer from drifts and instabilities. 

Rayleigh's curse is not integral to the problem, but rather an artifact of only considering the intensity of the field. By optimizing over all possible quantum measurements via the quantum Fisher information~\cite{Petz:2011aa}, it can be shown that the precision of an optimal measurement maintains a fairly constant value for any pulse separation $\tau$. In other words, the divergence can be averted using phase-sensitive measurements, despite the incoherent nature of the sources. This information is always available, no matter how small $\tau$ becomes. Experiments projecting onto tailored optical field modes have demonstrated considerably better precision than the direct-detection CRLB in both the spatial~\cite{Paur:2016aa,Tham:2016aa,Yang:2016aa} and the time domain~\cite{Donohue:2018aa}.
\begin{figure*}[t]
  \begin{center}
    \includegraphics[width=1\linewidth]{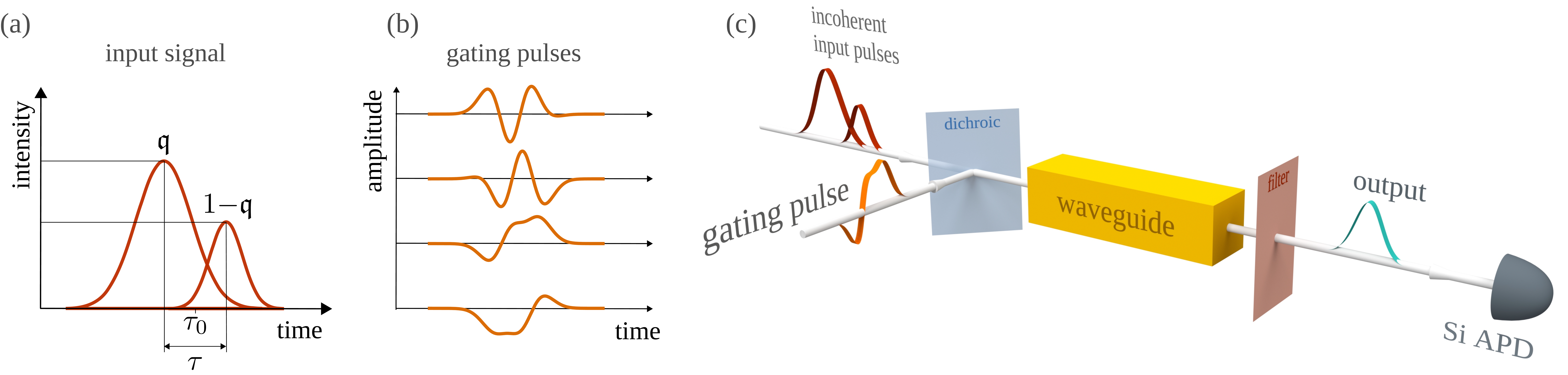}
  \end{center}
  \caption{Experimental concept. Two mutually incoherent light pulses can be characterized by a temporal separation, $\tau$, a joint temporal centroid position, $\tau_{0}$, and imbalanced intensities parametrized by $\mathfrak{q}$. The task is to find a measurement that facilitates the simultaneous estimation of all three parameters with the best possible precision. The right bottom panel shows the temporal envelopes of the projections used for our optimal estimation.}
  \label{fig:outline}
\end{figure*}

These results are as interesting as they are important, but they apply exclusively to signals of equal strength. Here, we consider a more broadly applicable multiparameter scenario in which the pulses might have different intensities. This can occur whenever an incoherently backscattered echo pulse is measured relative to a reference, e.g. in lidar ranging applications. This involves the simultaneous estimation of the temporal centroid, the time offset, and the relative intensities of the two pulses. Typically, when trying to estimate multiple parameters, there is a trade-off in the precision with which different parameters may be estimated; when the protocol is optimized for one parameter, its performance in estimating the remaining ones deteriorates. The underlying reason for this is an incompatibility of the quantum measurements required to simultaneously optimize the estimation of multiple parameters, meaning that it may not be possible to estimate all parameters optimally at the same time~\cite{Matsumoto:2002aa,Pezze:2017aa}. The theory of multiparameter estimation has attracted considerable interest during recent years, with promise for a variety of important applications~\cite{Szczykulska:2016aa,Albarelli:2020aa}.  To date, few recipes to saturate the ultimate precision bounds are known, and experimental demonstrations remain challenging~\cite{Vidrighin:2014aa,Polino:2019aa}.

In this work, we experimentally achieve the ultimate quantum limits for multiparameter timing estimation. We explicitly show that tailored strategies lead to a significant improvement in precision over direct detection for any number of photons. This constitutes not only a unique demonstration of multiparameter estimation at the quantum limit, but it works precisely in the regime in which classical detection entirely fails, thus solving the outstanding challenge of measuring extremely small time delays between faint, mutually incoherent pulses.

It is worth stressing that our method is clearly distinct from existing approaches that use quantum resources, such as squeezing or entanglement, to achieve better scaling of the measurement precision with respect to the number of photons~\cite{Giovannetti:2004aa}. These approaches rely on highly fragile probes, which are often not compatible with real-life conditions, such as strong losses within the system. In contrast, our approach focuses on performing an ideal measurement, making it versatile in both real-life applications and under extreme conditions, such as the faint-light astronomical measurements.

\section{Optimal measurements}

In the following, we briefly lay out the theory underlying our approach. A schematic setup is depicted in Fig~\ref{fig:outline}(a). Two pulses of identical amplitude shape $\psi ( t ) $, but different intensities, overlap with a time offset $\tau$ between them. In the case of direct detection, the signal acquired by a detector with perfect temporal resolution is 
\begin{equation}  
I ( t ) =  
\mathfrak{q} \; | \psi ( t ) |^{2}  + 
(1- \mathfrak{q}) \; | \psi (t - \tau) |^2 \, , 
\label{eq:signal}
\end{equation} 
where $\mathfrak{q}$ is the imbalance parameter $\mathfrak{q}$ that accounts for the different intensities. Note that we have assumed an incoherent mixture of the pulses~\cite{Watts:2017aa}, which is a good model for the common situation in which, e.g., one of the pulses is produced from the reflection of an incoherent scatterer. In our experiment, this incoherent signal is created by mixing the measurement outcomes of positively and negatively shifted pulses in time. To be specific, we focus on the case of Gaussian pulses, with temporal amplitude 
\begin{equation}
\psi (t \pm \tau/2) = 
\frac{1}{(2\pi\sigma_{\tau}^2)^\frac{1}{4}} 
\exp\left[-\frac{\left(t -\tau_0 \pm \tau/2\right)^2}
{4\sigma_\tau^2}\right] \, ,
\label{eq:gaussian}
\end{equation}
$\sigma_{\tau}$ being the root-mean-square (RMS) width.

Our goal is the simultaneous estimation of $\tau$, $\mathfrak{q}$ and the centroid $\tau_{0}$. The solution to this problem is to use phase-sensitive projections onto specifically designed temporal modes that are sketched in Fig.~\ref{fig:outline}(b). An optimal measurement basis is always given by the successive derivatives of the amplitude pulse shape {\cite{Rehacek:2017aa,Rehacek:2018aa}}: in our case, this basis is just the Hermite-Gauss modes $\{ \mathrm{HG}_{n}\}$. Following the procedure outlined in Ref.~\cite{Rehacek:2017aa}, it is enough with four mode projections $n=0, \ldots, 3$, which turn out to be
\begin{equation}
\begin{split}
\ket{\pi_0}=\begin{pmatrix}0\\ 1/\sqrt{6}\\ 1/\sqrt{2}\\ -1/\sqrt{3} \end{pmatrix},
\qquad
\ket{\pi_1}=\begin{pmatrix}0\\ 1/\sqrt{6}\\ -1/\sqrt{2}\\ -1/\sqrt{3} \end{pmatrix}, 
\quad \\
\\
\ket{\pi_2}=\begin{pmatrix}\sqrt{2}/\sqrt{5}\\ \sqrt{2}/\sqrt{5}\\ 0\\ 1/\sqrt{5} \end{pmatrix},
\qquad
\ket{\pi_3}=\begin{pmatrix}\sqrt{-3}/\sqrt{5}\\ 2/\sqrt{15}\\ 0\\ \sqrt{2}/\sqrt{15} \end{pmatrix}.
\end{split}
\end{equation}

To perform such projective measurements we use a quantum pulse gate \cite{Eckstein:2011aa, brecht2014demonstration, ansari2018tomography}, which is a reconfigurable temporal-mode demultiplexer shown in Fig.~\ref{fig:outline}(c). Gating pulses propagate through an optically nonlinear waveguide and interact with the incoherent near-infrared pulses to create an output signal at green wavelengths. Detecting the output photons correspond to a projective measurement along a certain temporal-mode set by the gating pulse. This method combines the advantage of the enhanced precision of phase-sensitive measurements with the simplicity and efficiency of photon counting at visible wavelengths.

\section{Experimental Setup}

\begin{figure}[t]
\begin{center}
    \includegraphics[width=1\columnwidth]{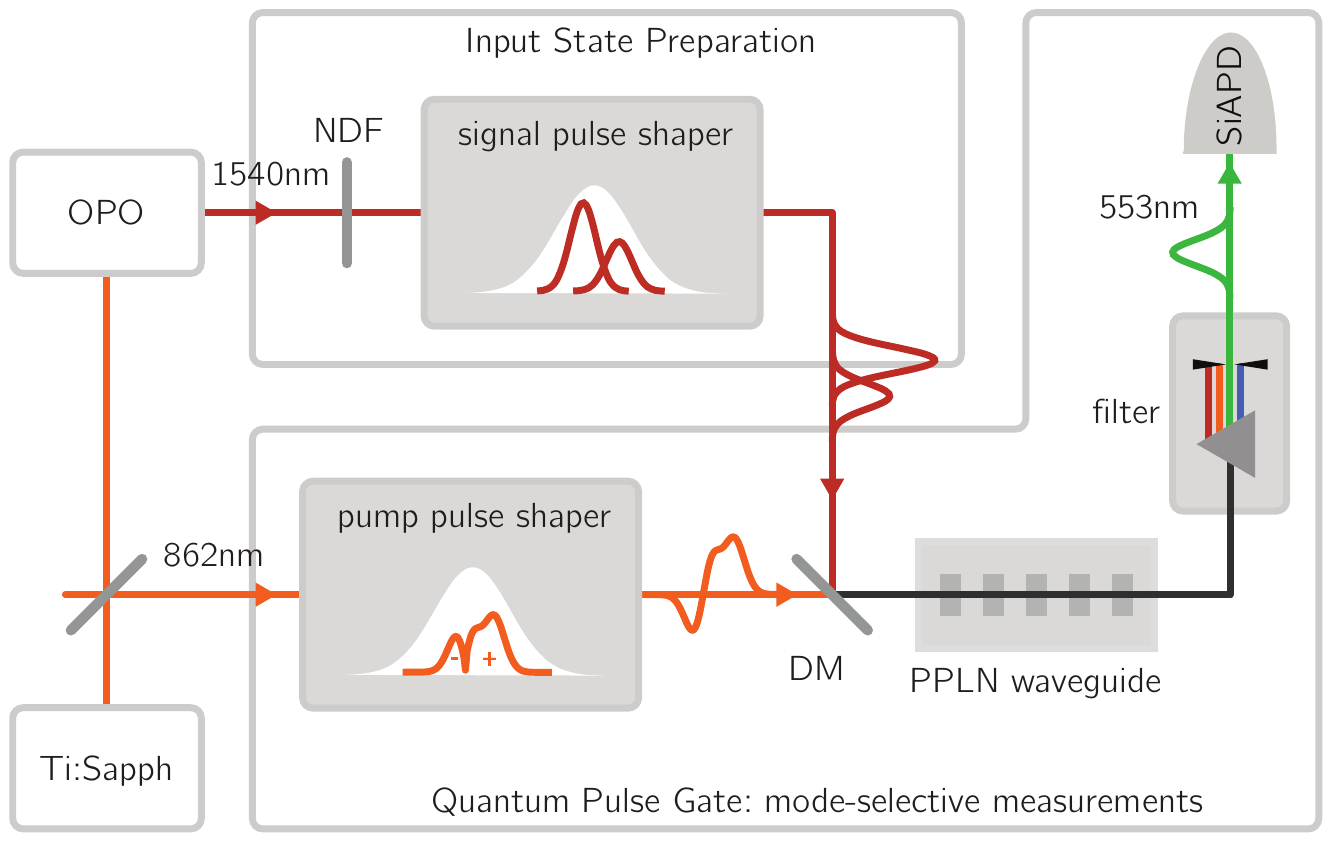}
  \end{center}
  \caption{Experimental setup. The input pulses with different time delays and intensities are carved from an attenuated broadband OPO at 1540~nm, using a Fourier-plane liquid-crystal spatial light modulator. Gating pulses with superpositions of Hermite-Gauss functions are shaped by a similar pulse shaper. We then mix the gating and input pulses in a PPLN waveguide, followed by bandpass filter and a silicon avalanche photo-diode (SiAPD) to count the up-converted green photons. Neutral density filter (NDF), dichroic mirror (DM).}
  \label{fig:expsetup}
\end{figure}

Our experimental apparatus is sketched in Fig.~\ref{fig:expsetup}. A titanium-sapphire oscillator and an optical parametric oscillator (OPO) are used to generate 150~fs long pulses at 862~nm and 1540~nm, respectively, with a repetition rate of 80~MHz. A commercial fiber-coupled pulse shaper carves two Gaussian pulses with durations of 1.57~ps from the OPO, one of which receives a positive time shift of $\tau/2$, whereas the other one receives a negative time shift of $-\tau/2$. These input pulses are also attenuated to a mean value of two photons per pulse. Ten pulse separations ranging from $0$ to $\tau$ and six imbalance parameter $\mathfrak{q}$ ranging from $0.125$ to $0.75$ were programmed during the experiment. Without loss of generality, we keep the centroid position, $\tau_{0}$, set at zero. Positively and negatively shifted pulses are measured separately, then an incoherent mixture of the two pulses, as in Eq.~(\ref{eq:signal}), is generated by mixing the individual measurement outcomes in data post-processing. This ensures that no spurious coherence can enter the measurements.  

The gating pulses, with a central wavelength of 862~nm, are shaped into the measurement modes, using a free space Fourier-plane spatial light modulator. The input pulses and the gating pulses are then sent to a quantum pulse gate~\cite{Eckstein:2011aa, brecht2014demonstration, ansari2018tomography}. As mentioned, this device is a mode-selective frequency converter which facilitates projections onto arbitrary temporal modes that are user-chosen and defined by the temporal-mode of the gating pulses.

Our quantum pulse gate is a 35~mm long titanium-indiffused periodically poled lithium niobate (PPLN) waveguide with a poling period of 4.4~$\mu$m. Propagation of the optical fields in the fundamental spatial mode of the waveguide is assured by the waveguide geometry for the input signal and by optical mode matching for the gating field. The gating pulses have a pulse energy of 100 pJ which provides a conversion efficiency of 40\%, excluding collection and detection losses. The sum-frequency generated light at 553~nm is filtered with a $4f$ setup to discard the phasematching sidelobes with a tight bandwidth of 17~GHz and then coupled to a single-mode fiber and detected with an off-the-shelf silicon avalanche photodiode. 

In our current quantum pulse gate implementation, the four modes are measured sequentially. This is not a fundamental limitation of the device; with a multiplexing scheme one can measure many modes in parallel, which can enable a single-shot multiparameter estimation~\cite{Silver:2019aa}. To collect statistics, each setting of input pulses and gating pulses was measured 100 times, with a total measurement time of 2 microsecond per setting. As our experiment is very sensitive to the smallest temporal drifts between input signal and gating field---after all, it was designed for exactly this purpose---special care was taken to limit temperature fluctuations of the measurement setup to below $0.1^{\circ}$~C.

\begin{figure}[t]
  \begin{center}
\begin{center}
    \includegraphics[width=1\linewidth]{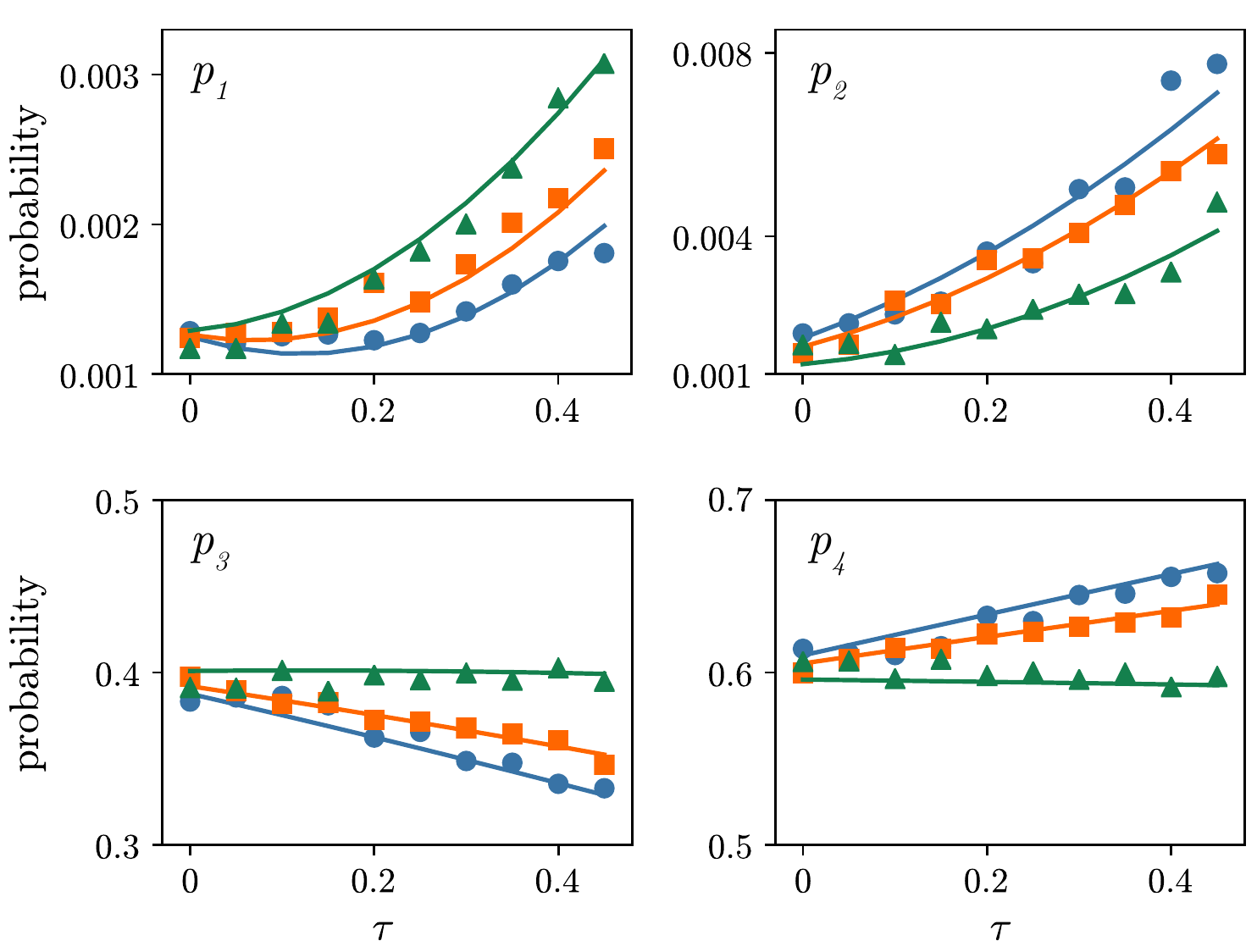}
  \end{center}
  \end{center}
  \caption{Measurement model. The panels show the mean measurement responses of our four measurement channels for a selected range of parameters. In all panels $\tau_0=0$ and $\mathfrak{q}=0.125$ (blue dots), $\mathfrak{q}=0.25$ (orange squares), and $\mathfrak{q}=0.5$ (green triangles). The curves show the corresponding responses of the theoretical detection model   (\ref{eq:meas_model}) used for multiparameter estimation.}
  \label{fig:measmod}
\end{figure}

\begin{figure*}[t]
  \begin{center}
 \includegraphics[width=.77\linewidth]{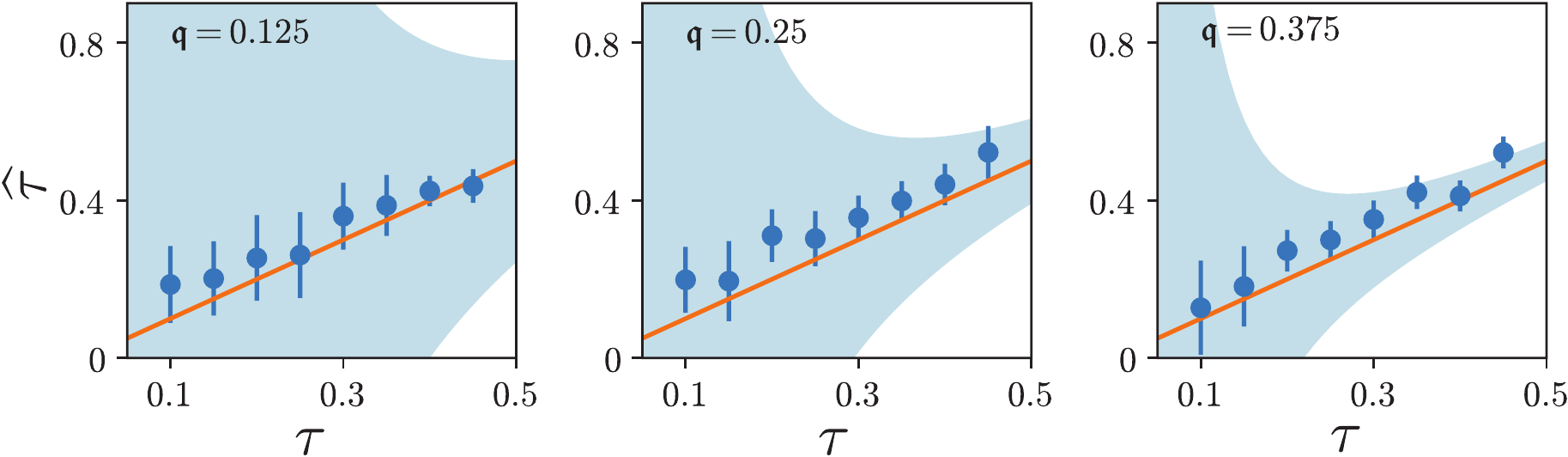}\\
 \includegraphics[width=.77\linewidth]{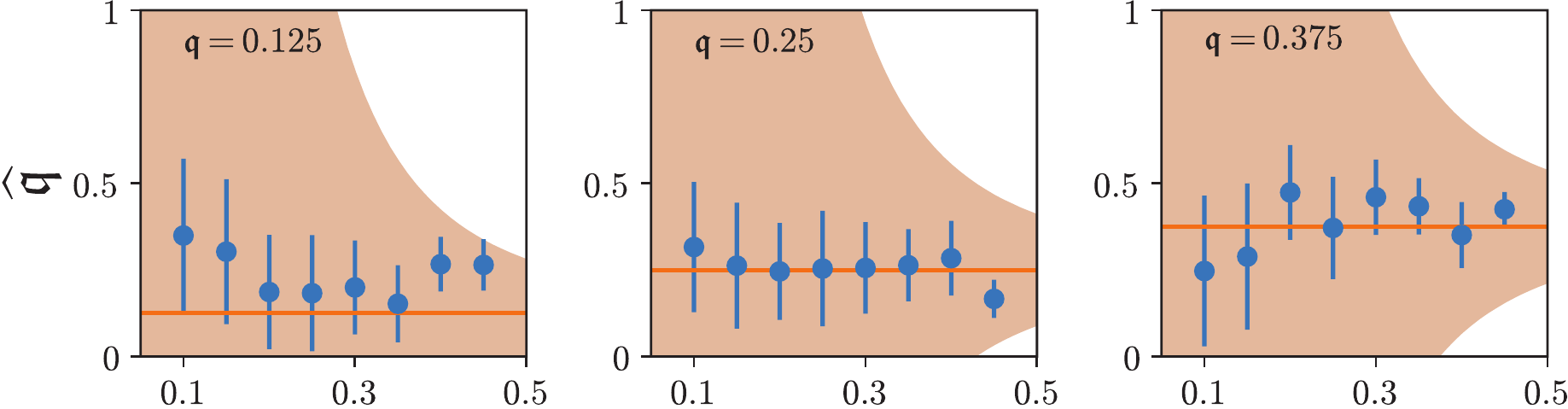}\\
 \includegraphics[width=.77\linewidth]{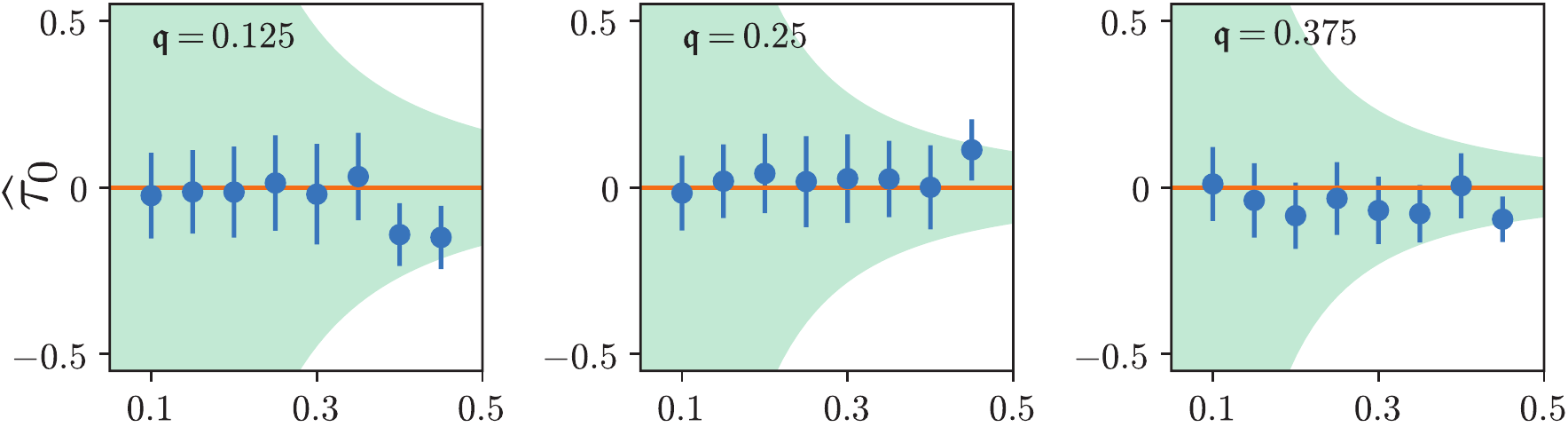}
  \end{center}
  \caption{Experimental results for the simultaneous estimation of the time offset $\widehat{\tau}$, the relative intensities $\widehat{\mathfrak{q}}$ and the temporal centroid $\widehat{\tau}_{0}$ of the incoherent mixture of two Gaussian pulses. Dimensionless time is obtained by scaling the time by the pulse  width $\sigma_{t}$.  Orange lines show true values of the corresponding measured parameters. Blue dots and error bars show the sample means and standard deviations over 100 estimates, each based on about 69000 detections. The shaded areas in all the plots represent the Cram\'{e}r-Rao lower bounds for direct intensity measurements corresponding to the same total number of photon counts.}
  \label{fig:est}
\end{figure*}

To construct an unbiased estimator resilient to the imperfect selectivity of our device, we use calibration data to perform measurement tomography of our technique~\cite{Rehacek:2017ab}. In particular, approximating the measurement responses; i.e., the  probabilities $p_j(\tau, \mathfrak{q}, \tau_0,)$ ($j=1,\ldots,4$) of the four implemented projective measurements, by low-order polynomials  quantities $\tau$ and $\tau_0$,
\begin{align}
\label{eq:meas_model}
p_{j}  = &  c_{0j} + c_{1j} \tau_{0}  + c_{2j} \tau + 
c_{3j} \mathfrak{q} + c_{4j} \tau_0^2 + 
c_{5j} \tau_0 \tau \nonumber \\
+ & c_{6j} \tau_0 \mathfrak{q}
+ c_{7j} \tau^2 + c_{8j} \tau \mathfrak{q} + 
c_{9j} \tau_0 \tau\mathfrak{q} \, ,
\end{align}
the unknown coefficients $c_{\alpha j}$ are estimated from data averaged over 100 repetitions using the generalized least squares (GLS) estimator from about 23 million total detections. This provides a theoretical description of the measurement apparatus and confirms that targeted optimality conditions, namely the (nearly) zero overlap of two of the measurement channels with the fundamental Gaussian mode, are obeyed by the laboratory setup.

Our measurement model is used, in turn, for constructing multiparameter estimates from individual measurement runs, each comprising about 23 thousands of detections. In this case, constrained GLS estimation is applied for inverting the nonlinear system \eqref{eq:meas_model} to ensure that the physical constraints $\widehat{\tau}\ge 0$ and $0 \le \widehat{\mathfrak{q}}\le 1$ are obeyed by the estimates. As nonnegativity of $\tau$ makes the corresponding estimator $\widehat{\tau}$ biased for very small separations, slight violations of the quantum CRLB might be seen in such extreme cases. 

The results are shown in Fig.~\ref{fig:measmod}, which presents the responses of our four measurement channels.  Notice the realized measurement has two ``dark'' channels with almost no intensity in the limit $\tau \rightarrow 0$, as required by optimality criteria granting superior performance of the quantum measurement over best intensity based inference.

\section{Results and Discussion}

In Fig.~\ref{fig:est} we present our experimental results for the the simultaneous estimation of the three parameters: time delay ($\widehat{\tau}$), intensity imbalance ($\widehat{\mathfrak{q}}$) and time-delay centroid ($\widehat{\tau}_{0}$). The solid orange lines are the programmed true values of the variables, whereas the shaded regions mark the precision limits of incoherent direct detection, as derived directly from the CRLB. 

For very small time separations, no meaningful information can be extracted from direct measurements, particularly in the case of strong imbalance between the intensities of the two signal pulses. The blue dots are the estimates retrieved from our measurements. The corresponding errors are computed from maximum likelihood. 

\begin{figure}[t]
  \begin{center}
 \includegraphics[width=.7\columnwidth]{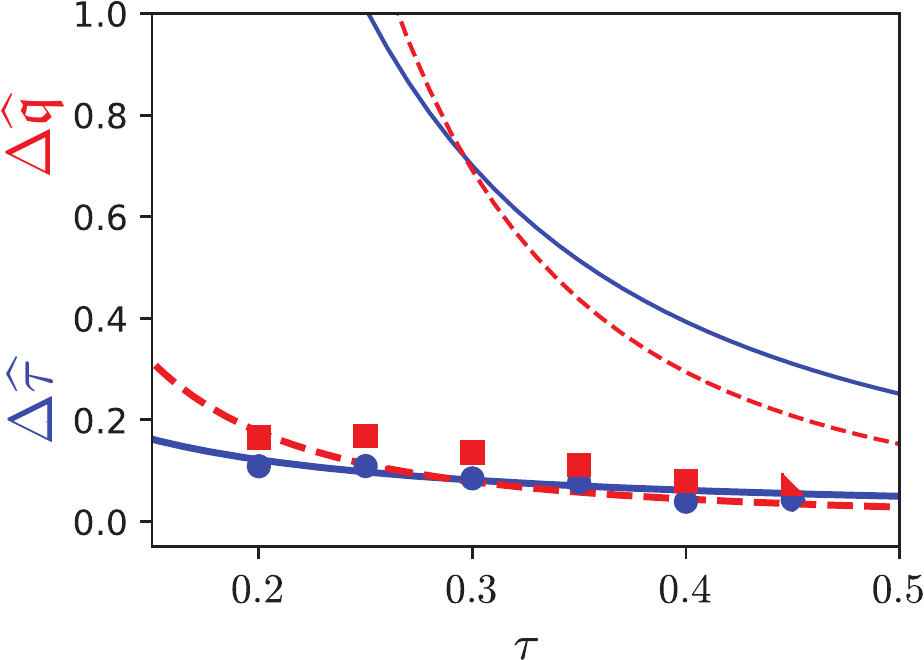}
   \end{center}
  \caption{Precision of the method. Variances of the estimator $\widehat{\tau}$ (blue circles) and $\widehat{\mathfrak{q}}$ (red squares) from our experimental data. Thick lines give the quantum CRLB, and thin lines correspond to the direct-detection CRLB. The data corresponds to $\mathfrak{q} = 0.125$ and 69000 effective detections.}
  \label{fig:qlimest}
\end{figure}


In Fig.~\ref{fig:qlimest} we plot the variances of the estimators $\widehat{\tau}$ and $\widehat{\mathfrak{q}}$ from our experimental data. A clear separation is seen between the direct-detection CRLB and the true quantum CRLB. The experimental data strongly outperforms direct-detection strategies and approaches the quantum CRLB for all measured separations and imbalance parameters, confirming that we have indeed implemented an ideal measurement that yields the maximum achievable information for this multiparameter estimation problem. We also note that, for some values of $\tau$, the variances are slightly below the CRLB: this is due to the systematic errors in producing the signal state. At very small true separations ($\tau \lesssim 0.2$ in our dimensionless units) the estimator becomes biased, and the CRLB must be adapted~\cite{Eldar:2006aa}.

These results demonstrate that mode-selective time  measurements in the proper optimal modes constitute a unique tool for precision parameter estimation problems where intensity measurements fail. Notably, the absolute time and frequency scales accessible are not strongly dependent on the scale of the measurement pulses, but rather on the exact implementation of the mode-sensitive detector~\cite{ansari2018tailoring}. In our realization, this corresponds to time and frequency scales of 30~fs and 17~GHz, respectively.

In summary, our results show that multiparameter estimation in the time-frequency domain can benefit greatly from quantum-inspired techniques and analysis. By exploiting time-frequency mode-selective measurements, we have shown that multiple parameters, including sub-pulse-width separations, relative intensities, and delay-pulse centroid, can be estimated simultaneously with precision below the standard CRLB. By adapting these techniques to different scales, this method could find immediate practical use in atomic and stellar spectral characterization and  time-of-flight imaging and spectroscopy.

\section{Acknowledgments}
We acknowledge financial support from the European Union's Horizon 2020 research and innovation program (Project ApresSF), the Deutsche Forschungsgemeinschaft (Grant 231447078--TRR 142), the Grant Agency of the Czech Republic (Grant 18-04291S), the Spanish Ministerio de Ciencia e Innovaci\'on (Grant PGC2018-099183-B-I00), and the Q-FARM Bloch Fellowship.

\section*{Appendix A: Ultimate limits of the standard measurements of ultrashort pulses}

In order to characterize the temporal shape of an ultrashort pulse it is often convenient to combine the pulse with itself. Varying the delay between the pulse copies and measuring the signal at each delay gives an estimate of the pulse duration.  These autocorrelation measurements have limitations: to estimate the  duration requires assuming a particular pulse shape, and the phase of the pulse electric field cannot be measured at all. 

A variety of methods have been devised to bypass these drawbacks. We will focus here to one of the most popular, the so-called frequency-resolved optical gating (FROG), although our analysis can be extended to other similar techniques. FROG and autocorrelation share the idea of combining a pulse with itself in a nonlinear medium. But FROG measures the spectrum of the signal at each delay $T$ (hence the term \emph{frequency-resolved}), instead of just the intensity. This measurement creates a spectrogram of the pulse; i.e.,  
\begin{equation}
I_{\mathrm{FROG}} (\omega, T) = | E_{\mathrm{sig}} (\omega, T ) |^{2} =
\left | \int_{- \infty}^{+ \infty} E_{\mathrm{sig}} (t, T ) e^{- i \omega t} \, dt \right |^{2} ,
\end{equation}
where $E_{\mathrm{sig}}$ is the signal field from the nonlinear interaction. This field depends on the original pulse and the nonlinear process employed. For the common case in which the second-harmonic generation is used, we have that $E_{\mathrm{sig}} (t, T ) = E(t) \, E(t - T)$, so that 
\begin{equation}
I_{\mathrm{SHG \; FROG}} (\omega, T) =  
\left | \int_{- \infty}^{+ \infty} E(t) \, E (t - T ) \; e^{- i \omega t} \, dt \right |^{2} \, ,
\end{equation}
 which can be used to determine the complex electric field as a function of time or frequency.

FROG is currently one of the most widespread techniques for measuring ultrashort laser pulses.  It allows for the use of a phase-retrieval algorithm to retrieve the precise pulse intensity and phase vs. time. 

Now, let us imagine that we use FROG for resolving the temporal separation $\tau$ between two incoherent pulses. To simplify the problem, we take both pulses to be of identical shape $E(t)$. The basic signal is then
\begin{equation}
E(t - \tau/2) + e^{i \phi} E ( t +\tau/2) \, ,
\end{equation}
where we have to average over $\phi$ to take into account that we are dealing with an incoherent mixture. The corresponding spectrogram reads
 \begin{equation}
 I (\omega, T, \tau) = 2 \cos (\tau T) \; I_{\mathrm{SHG \; FROG}} (\omega, T) 
 + A (\tau) \, A(- \tau ) \, ,
 \end{equation}
with
 \begin{equation}
 A (\tau ) = \int_{- \infty}^{+ \infty} E(t-\tau/2) \, E (t + \tau/2 - T) \, 
 e^{- i \omega t} \, dt  \, .
 \end{equation}
Since $\tau$ is the variable of interest, we write $I(\tau) \equiv I(\omega, T, \tau)$. The crucial observation for what follows is that the series expansion of $I(\tau)$ in terms of $\tau$ has no linear term; that is, the detected signal depends quadratically on $\tau$:
\begin{equation}
I  (\tau) = I  (0) + \tau^{2} I^{\prime \prime} (0) \, .
\end{equation}
We can then estimate the uncertainty in our measurement via simple linear error propagation; we get for the variances
\begin{equation}
\Delta^{2} \tau = \frac{\Delta^{2} I}{(\partial_{\tau} I)^{2}} \, .
\end{equation}
But, since  $\partial_{\tau} I$ goes to zero for small $\tau$, the uncertainty  diverges and the method does suffer from Rayleigh's curse. Hence, established pulse characterization methods fail when operated on incoherent pulse mixtures and other methods have to be used; e.g., the approach demonstrated in the main text.


\begin{thebibliography}{38}%
\makeatletter
\providecommand \@ifxundefined [1]{%
 \@ifx{#1\undefined}
}%
\providecommand \@ifnum [1]{%
 \ifnum #1\expandafter \@firstoftwo
 \else \expandafter \@secondoftwo
 \fi
}%
\providecommand \@ifx [1]{%
 \ifx #1\expandafter \@firstoftwo
 \else \expandafter \@secondoftwo
 \fi
}%
\providecommand \natexlab [1]{#1}%
\providecommand \enquote  [1]{``#1''}%
\providecommand \bibnamefont  [1]{#1}%
\providecommand \bibfnamefont [1]{#1}%
\providecommand \citenamefont [1]{#1}%
\providecommand \href@noop [0]{\@secondoftwo}%
\providecommand \href [0]{\begingroup \@sanitize@url \@href}%
\providecommand \@href[1]{\@@startlink{#1}\@@href}%
\providecommand \@@href[1]{\endgroup#1\@@endlink}%
\providecommand \@sanitize@url [0]{\catcode `\\12\catcode `\$12\catcode
  `\&12\catcode `\#12\catcode `\^12\catcode `\_12\catcode `\%12\relax}%
\providecommand \@@startlink[1]{}%
\providecommand \@@endlink[0]{}%
\providecommand \url  [0]{\begingroup\@sanitize@url \@url }%
\providecommand \@url [1]{\endgroup\@href {#1}{\urlprefix }}%
\providecommand \urlprefix  [0]{URL }%
\providecommand \Eprint [0]{\href }%
\providecommand \doibase [0]{http://dx.doi.org/}%
\providecommand \selectlanguage [0]{\@gobble}%
\providecommand \bibinfo  [0]{\@secondoftwo}%
\providecommand \bibfield  [0]{\@secondoftwo}%
\providecommand \translation [1]{[#1]}%
\providecommand \BibitemOpen [0]{}%
\providecommand \bibitemStop [0]{}%
\providecommand \bibitemNoStop [0]{.\EOS\space}%
\providecommand \EOS [0]{\spacefactor3000\relax}%
\providecommand \BibitemShut  [1]{\csname bibitem#1\endcsname}%
\let\auto@bib@innerbib\@empty
\bibitem [{\citenamefont {Droste}\ \emph {et~al.}(2015)\citenamefont {Droste},
  \citenamefont {Grebing}, \citenamefont {Leute}, \citenamefont {Raupach},
  \citenamefont {Matveev}, \citenamefont {H{\"a}nsch}, \citenamefont {Bauch},
  \citenamefont {Holzwarth},\ and\ \citenamefont {Grosche}}]{Droste:2015aa}%
  \BibitemOpen
  \bibfield  {author} {\bibinfo {author} {\bibfnamefont {S.}~\bibnamefont
  {Droste}}, \bibinfo {author} {\bibfnamefont {C.}~\bibnamefont {Grebing}},
  \bibinfo {author} {\bibfnamefont {J.}~\bibnamefont {Leute}}, \bibinfo
  {author} {\bibfnamefont {S.~M.}\ \bibnamefont {Raupach}}, \bibinfo {author}
  {\bibfnamefont {A.}~\bibnamefont {Matveev}}, \bibinfo {author} {\bibfnamefont
  {T.~W.}\ \bibnamefont {H{\"a}nsch}}, \bibinfo {author} {\bibfnamefont
  {A.}~\bibnamefont {Bauch}}, \bibinfo {author} {\bibfnamefont
  {R.}~\bibnamefont {Holzwarth}}, \ and\ \bibinfo {author} {\bibfnamefont
  {G.}~\bibnamefont {Grosche}},\ }\href@noop {} {\bibfield  {journal} {\bibinfo
   {journal} {New Journal of Physics}\ }\textbf {\bibinfo {volume} {17}},\
  \bibinfo {pages} {083044} (\bibinfo {year} {2015})}\BibitemShut {NoStop}%
\bibitem [{\citenamefont {Chou}\ \emph {et~al.}(2010)\citenamefont {Chou},
  \citenamefont {Hume}, \citenamefont {Rosenband},\ and\ \citenamefont
  {Wineland}}]{Chou:2010aa}%
  \BibitemOpen
  \bibfield  {author} {\bibinfo {author} {\bibfnamefont {C.~W.}\ \bibnamefont
  {Chou}}, \bibinfo {author} {\bibfnamefont {D.~B.}\ \bibnamefont {Hume}},
  \bibinfo {author} {\bibfnamefont {T.}~\bibnamefont {Rosenband}}, \ and\
  \bibinfo {author} {\bibfnamefont {D.~J.}\ \bibnamefont {Wineland}},\ }\href
  {\doibase 10.1126/science.1192720} {\bibfield  {journal} {\bibinfo  {journal}
  {Science}\ }\textbf {\bibinfo {volume} {329}},\ \bibinfo {pages} {1630}
  (\bibinfo {year} {2010})}\BibitemShut {NoStop}%
\bibitem [{\citenamefont {Krehlik}\ \emph {et~al.}(2017)\citenamefont
  {Krehlik}, \citenamefont {Buczek}, \citenamefont {Ko{\l}odziej},
  \citenamefont {Lipi{\'n}ski}, \citenamefont {{\'S}liwczy{\'n}ski},
  \citenamefont {Nawrocki}, \citenamefont {Noga{\'s}}, \citenamefont {Marecki},
  \citenamefont {Pazderski}, \citenamefont {Ablewski}, \citenamefont {Bober},
  \citenamefont {Ciury{\l}o}, \citenamefont {Cygan}, \citenamefont {Lisak},
  \citenamefont {Mas{\l}owski}, \citenamefont {Morzy{\'n}ski}, \citenamefont
  {Zawada}, \citenamefont {Campbell}, \citenamefont {Pieczerak}, \citenamefont
  {Binczewski},\ and\ \citenamefont {Turza}}]{Krehlik:2017aa}%
  \BibitemOpen
  \bibfield  {author} {\bibinfo {author} {\bibfnamefont {P.}~\bibnamefont
  {Krehlik}}, \bibinfo {author} {\bibfnamefont {{\L}.}~\bibnamefont {Buczek}},
  \bibinfo {author} {\bibfnamefont {J.}~\bibnamefont {Ko{\l}odziej}}, \bibinfo
  {author} {\bibfnamefont {M.}~\bibnamefont {Lipi{\'n}ski}}, \bibinfo {author}
  {\bibfnamefont {{\L}.}~\bibnamefont {{\'S}liwczy{\'n}ski}}, \bibinfo {author}
  {\bibfnamefont {J.}~\bibnamefont {Nawrocki}}, \bibinfo {author}
  {\bibfnamefont {P.}~\bibnamefont {Noga{\'s}}}, \bibinfo {author}
  {\bibfnamefont {A.}~\bibnamefont {Marecki}}, \bibinfo {author} {\bibfnamefont
  {E.}~\bibnamefont {Pazderski}}, \bibinfo {author} {\bibfnamefont
  {P.}~\bibnamefont {Ablewski}}, \bibinfo {author} {\bibfnamefont
  {M.}~\bibnamefont {Bober}}, \bibinfo {author} {\bibfnamefont
  {R.}~\bibnamefont {Ciury{\l}o}}, \bibinfo {author} {\bibfnamefont
  {A.}~\bibnamefont {Cygan}}, \bibinfo {author} {\bibfnamefont
  {D.}~\bibnamefont {Lisak}}, \bibinfo {author} {\bibfnamefont
  {P.}~\bibnamefont {Mas{\l}owski}}, \bibinfo {author} {\bibfnamefont
  {P.}~\bibnamefont {Morzy{\'n}ski}}, \bibinfo {author} {\bibfnamefont
  {M.}~\bibnamefont {Zawada}}, \bibinfo {author} {\bibfnamefont {R.~M.}\
  \bibnamefont {Campbell}}, \bibinfo {author} {\bibfnamefont {J.}~\bibnamefont
  {Pieczerak}}, \bibinfo {author} {\bibfnamefont {A.}~\bibnamefont
  {Binczewski}}, \ and\ \bibinfo {author} {\bibfnamefont {K.}~\bibnamefont
  {Turza}},\ }\href {https://doi.org/10.1051/0004-6361/201730615} {\bibfield
  {journal} {\bibinfo  {journal} {A\&A}\ }\textbf {\bibinfo {volume} {603}},\
  \bibinfo {pages} {A48} (\bibinfo {year} {2017})}\BibitemShut {NoStop}%
\bibitem [{\citenamefont {Fujimoto}\ \emph {et~al.}(1995)\citenamefont
  {Fujimoto}, \citenamefont {Brezinski}, \citenamefont {Tearney}, \citenamefont
  {Boppart}, \citenamefont {Bouma}, \citenamefont {Hee}, \citenamefont
  {Southern},\ and\ \citenamefont {Swanson}}]{Fujimoto:1995aa}%
  \BibitemOpen
  \bibfield  {author} {\bibinfo {author} {\bibfnamefont {J.~G.}\ \bibnamefont
  {Fujimoto}}, \bibinfo {author} {\bibfnamefont {M.~E.}\ \bibnamefont
  {Brezinski}}, \bibinfo {author} {\bibfnamefont {G.~J.}\ \bibnamefont
  {Tearney}}, \bibinfo {author} {\bibfnamefont {S.~A.}\ \bibnamefont
  {Boppart}}, \bibinfo {author} {\bibfnamefont {B.}~\bibnamefont {Bouma}},
  \bibinfo {author} {\bibfnamefont {M.~R.}\ \bibnamefont {Hee}}, \bibinfo
  {author} {\bibfnamefont {J.~F.}\ \bibnamefont {Southern}}, \ and\ \bibinfo
  {author} {\bibfnamefont {E.~A.}\ \bibnamefont {Swanson}},\ }\href {\doibase
  10.1038/nm0995-970} {\bibfield  {journal} {\bibinfo  {journal} {Nat. Med.}\
  }\textbf {\bibinfo {volume} {1}},\ \bibinfo {pages} {970} (\bibinfo {year}
  {1995})}\BibitemShut {NoStop}%
\bibitem [{\citenamefont {Graham}\ \emph {et~al.}(2013)\citenamefont {Graham},
  \citenamefont {Hogan}, \citenamefont {Kasevich},\ and\ \citenamefont
  {Rajendran}}]{Graham:2013aa}%
  \BibitemOpen
  \bibfield  {author} {\bibinfo {author} {\bibfnamefont {P.~W.}\ \bibnamefont
  {Graham}}, \bibinfo {author} {\bibfnamefont {J.~M.}\ \bibnamefont {Hogan}},
  \bibinfo {author} {\bibfnamefont {M.~A.}\ \bibnamefont {Kasevich}}, \ and\
  \bibinfo {author} {\bibfnamefont {S.}~\bibnamefont {Rajendran}},\ }\href
  {\doibase 10.1103/PhysRevLett.110.171102} {\bibfield  {journal} {\bibinfo
  {journal} {Phys. Rev. Lett.}\ }\textbf {\bibinfo {volume} {110}},\ \bibinfo
  {pages} {171102} (\bibinfo {year} {2013})}\BibitemShut {NoStop}%
\bibitem [{\citenamefont {Weiner}(2009)}]{Weiner:2009aa}%
  \BibitemOpen
  \bibfield  {author} {\bibinfo {author} {\bibfnamefont {A.}~\bibnamefont
  {Weiner}},\ }\href@noop {} {\emph {\bibinfo {title} {Ultrafast {O}ptics}}}\
  (\bibinfo  {publisher} {Wiley},\ \bibinfo {address} {Hoboken, NJ},\ \bibinfo
  {year} {2009})\BibitemShut {NoStop}%
\bibitem [{\citenamefont {Hansard}\ \emph {et~al.}(2013)\citenamefont
  {Hansard}, \citenamefont {Lee}, \citenamefont {Choi},\ and\ \citenamefont
  {Horaud}}]{Hansard:2013aa}%
  \BibitemOpen
  \bibfield  {author} {\bibinfo {author} {\bibfnamefont {M.}~\bibnamefont
  {Hansard}}, \bibinfo {author} {\bibfnamefont {S.}~\bibnamefont {Lee}},
  \bibinfo {author} {\bibfnamefont {O.}~\bibnamefont {Choi}}, \ and\ \bibinfo
  {author} {\bibfnamefont {R.}~\bibnamefont {Horaud}},\ }\href@noop {} {\emph
  {\bibinfo {title} {Time-of-Flight Cameras}}}\ (\bibinfo  {publisher}
  {Springer},\ \bibinfo {address} {Berlin},\ \bibinfo {year}
  {2013})\BibitemShut {NoStop}%
\bibitem [{\citenamefont {Maddaloni}\ \emph {et~al.}(2013)\citenamefont
  {Maddaloni}, \citenamefont {Bellini},\ and\ \citenamefont
  {Natale}}]{Maddaloni:2013aa}%
  \BibitemOpen
  \bibfield  {author} {\bibinfo {author} {\bibfnamefont {P.}~\bibnamefont
  {Maddaloni}}, \bibinfo {author} {\bibfnamefont {M.}~\bibnamefont {Bellini}},
  \ and\ \bibinfo {author} {\bibfnamefont {P.~D.}\ \bibnamefont {Natale}},\
  }\href@noop {} {\emph {\bibinfo {title} {Laser-based {M}easurements for
  {T}ime and {F}requency {D}omain {A}pplications}}}\ (\bibinfo  {publisher}
  {CRC},\ \bibinfo {address} {Boca Raton, FL},\ \bibinfo {year}
  {2013})\BibitemShut {NoStop}%
\bibitem [{\citenamefont {Trebino}(2002)}]{Trebino:2002aa}%
  \BibitemOpen
  \bibfield  {author} {\bibinfo {author} {\bibfnamefont {R.}~\bibnamefont
  {Trebino}},\ }\href@noop {} {\emph {\bibinfo {title} {Frequency-{R}esolved
  {O}ptical {G}ating}}}\ (\bibinfo  {publisher} {Springer},\ \bibinfo {address}
  {New York},\ \bibinfo {year} {2002})\BibitemShut {NoStop}%
\bibitem [{\citenamefont {Walmsley}\ and\ \citenamefont
  {Dorrer}(2009)}]{walmsley2009characterization}%
  \BibitemOpen
  \bibfield  {author} {\bibinfo {author} {\bibfnamefont {I.~A.}\ \bibnamefont
  {Walmsley}}\ and\ \bibinfo {author} {\bibfnamefont {C.}~\bibnamefont
  {Dorrer}},\ }\href@noop {} {\bibfield  {journal} {\bibinfo  {journal}
  {Advances in Optics and Photonics}\ }\textbf {\bibinfo {volume} {1}},\
  \bibinfo {pages} {308} (\bibinfo {year} {2009})}\BibitemShut {NoStop}%
\bibitem [{\citenamefont {Chang}(2011)}]{Chang:2011aa}%
  \BibitemOpen
  \bibfield  {author} {\bibinfo {author} {\bibfnamefont {Z.}~\bibnamefont
  {Chang}},\ }\href@noop {} {\emph {\bibinfo {title} {Fundamentals of
  {A}ttosecond {O}ptics}}}\ (\bibinfo  {publisher} {CRC},\ \bibinfo {address}
  {Boca Raton, FL},\ \bibinfo {year} {2011})\BibitemShut {NoStop}%
\bibitem [{\citenamefont {Nuss}\ \emph {et~al.}(1994)\citenamefont {Nuss},
  \citenamefont {Li}, \citenamefont {Chiu}, \citenamefont {Weiner},\ and\
  \citenamefont {Partovi}}]{Nuss:1994aa}%
  \BibitemOpen
  \bibfield  {author} {\bibinfo {author} {\bibfnamefont {M.~C.}\ \bibnamefont
  {Nuss}}, \bibinfo {author} {\bibfnamefont {M.}~\bibnamefont {Li}}, \bibinfo
  {author} {\bibfnamefont {T.~H.}\ \bibnamefont {Chiu}}, \bibinfo {author}
  {\bibfnamefont {A.~M.}\ \bibnamefont {Weiner}}, \ and\ \bibinfo {author}
  {\bibfnamefont {A.}~\bibnamefont {Partovi}},\ }\href {\doibase
  10.1364/OL.19.000664} {\bibfield  {journal} {\bibinfo  {journal} {Opt.
  Lett.}\ }\textbf {\bibinfo {volume} {19}},\ \bibinfo {pages} {664} (\bibinfo
  {year} {1994})}\BibitemShut {NoStop}%
\bibitem [{\citenamefont {Tsang}\ \emph {et~al.}(2016)\citenamefont {Tsang},
  \citenamefont {Nair},\ and\ \citenamefont {Lu}}]{Tsang:2016aa}%
  \BibitemOpen
  \bibfield  {author} {\bibinfo {author} {\bibfnamefont {M.}~\bibnamefont
  {Tsang}}, \bibinfo {author} {\bibfnamefont {R.}~\bibnamefont {Nair}}, \ and\
  \bibinfo {author} {\bibfnamefont {X.-M.}\ \bibnamefont {Lu}},\ }\href
  {http://link.aps.org/doi/10.1103/PhysRevX.6.031033} {\bibfield  {journal}
  {\bibinfo  {journal} {Phys. Rev. X}\ }\textbf {\bibinfo {volume} {6}},\
  \bibinfo {pages} {031033} (\bibinfo {year} {2016})}\BibitemShut {NoStop}%
\bibitem [{\citenamefont {Donohue}\ \emph {et~al.}(2018)\citenamefont
  {Donohue}, \citenamefont {Ansari}, \citenamefont {{\v R}eh{\'a}{\v c}ek},
  \citenamefont {Hradil}, \citenamefont {Stoklasa}, \citenamefont {Pa{\'u}r},
  \citenamefont {S{\'a}nchez-Soto},\ and\ \citenamefont
  {Silberhorn}}]{Donohue:2018aa}%
  \BibitemOpen
  \bibfield  {author} {\bibinfo {author} {\bibfnamefont {J.~M.}\ \bibnamefont
  {Donohue}}, \bibinfo {author} {\bibfnamefont {V.}~\bibnamefont {Ansari}},
  \bibinfo {author} {\bibfnamefont {J.}~\bibnamefont {{\v R}eh{\'a}{\v c}ek}},
  \bibinfo {author} {\bibfnamefont {Z.}~\bibnamefont {Hradil}}, \bibinfo
  {author} {\bibfnamefont {B.}~\bibnamefont {Stoklasa}}, \bibinfo {author}
  {\bibfnamefont {M.}~\bibnamefont {Pa{\'u}r}}, \bibinfo {author}
  {\bibfnamefont {L.~L.}\ \bibnamefont {S{\'a}nchez-Soto}}, \ and\ \bibinfo
  {author} {\bibfnamefont {C.}~\bibnamefont {Silberhorn}},\ }\href {\doibase
  10.1103/PhysRevLett.121.090501} {\bibfield  {journal} {\bibinfo  {journal}
  {Phys. Rev. Lett.}\ }\textbf {\bibinfo {volume} {121}},\ \bibinfo {pages}
  {090501} (\bibinfo {year} {2018})}\BibitemShut {NoStop}%
\bibitem [{\citenamefont {Kay}(1993)}]{Kay:1993aa}%
  \BibitemOpen
  \bibfield  {author} {\bibinfo {author} {\bibfnamefont {S.~M.}\ \bibnamefont
  {Kay}},\ }\href@noop {} {\emph {\bibinfo {title} {Fundamentals of Statistical
  Signal Processing---Estimation Theory}}},\ Vol.~\bibinfo {volume} {I}\
  (\bibinfo  {publisher} {Prentice Hall},\ \bibinfo {address} {New York},\
  \bibinfo {year} {1993})\BibitemShut {NoStop}%
\bibitem [{\citenamefont {Lamine}\ \emph {et~al.}(2008)\citenamefont {Lamine},
  \citenamefont {Fabre},\ and\ \citenamefont {Treps}}]{Lamine:2008aa}%
  \BibitemOpen
  \bibfield  {author} {\bibinfo {author} {\bibfnamefont {B.}~\bibnamefont
  {Lamine}}, \bibinfo {author} {\bibfnamefont {C.}~\bibnamefont {Fabre}}, \
  and\ \bibinfo {author} {\bibfnamefont {N.}~\bibnamefont {Treps}},\ }\href
  {https://link.aps.org/doi/10.1103/PhysRevLett.101.123601} {\bibfield
  {journal} {\bibinfo  {journal} {Phys. Rev. Lett.}\ }\textbf {\bibinfo
  {volume} {101}},\ \bibinfo {pages} {123601} (\bibinfo {year}
  {2008})}\BibitemShut {NoStop}%
\bibitem [{\citenamefont {Caves}\ \emph {et~al.}(1980)\citenamefont {Caves},
  \citenamefont {Thorne}, \citenamefont {Drever}, \citenamefont {Sandberg},\
  and\ \citenamefont {Zimmermann}}]{Caves:1980aa}%
  \BibitemOpen
  \bibfield  {author} {\bibinfo {author} {\bibfnamefont {C.~M.}\ \bibnamefont
  {Caves}}, \bibinfo {author} {\bibfnamefont {K.~S.}\ \bibnamefont {Thorne}},
  \bibinfo {author} {\bibfnamefont {R.~W.~P.}\ \bibnamefont {Drever}}, \bibinfo
  {author} {\bibfnamefont {V.~D.}\ \bibnamefont {Sandberg}}, \ and\ \bibinfo
  {author} {\bibfnamefont {M.}~\bibnamefont {Zimmermann}},\ }\href {\doibase
  10.1103/RevModPhys.52.341} {\bibfield  {journal} {\bibinfo  {journal} {Rev.
  Mod. Phys.}\ }\textbf {\bibinfo {volume} {52}},\ \bibinfo {pages} {341}
  (\bibinfo {year} {1980})}\BibitemShut {NoStop}%
\bibitem [{\citenamefont {Giovannetti}\ \emph {et~al.}(2004)\citenamefont
  {Giovannetti}, \citenamefont {Lloyd},\ and\ \citenamefont
  {Maccone}}]{Giovannetti:2004aa}%
  \BibitemOpen
  \bibfield  {author} {\bibinfo {author} {\bibfnamefont {V.}~\bibnamefont
  {Giovannetti}}, \bibinfo {author} {\bibfnamefont {S.}~\bibnamefont {Lloyd}},
  \ and\ \bibinfo {author} {\bibfnamefont {L.}~\bibnamefont {Maccone}},\ }\href
  {\doibase 10.1126/science.1104149} {\bibfield  {journal} {\bibinfo  {journal}
  {Science}\ }\textbf {\bibinfo {volume} {306}},\ \bibinfo {pages} {1330}
  (\bibinfo {year} {2004})}\BibitemShut {NoStop}%
\bibitem [{\citenamefont {Petz}\ and\ \citenamefont
  {Ghinea}(2011)}]{Petz:2011aa}%
  \BibitemOpen
  \bibfield  {author} {\bibinfo {author} {\bibfnamefont {D.}~\bibnamefont
  {Petz}}\ and\ \bibinfo {author} {\bibfnamefont {C.}~\bibnamefont {Ghinea}},\
  }\enquote {\bibinfo {title} {Introduction to {Q}uantum {F}isher
  {I}nformation},}\ in\ \href {\doibase doi:10.1142/9789814338745_0015} {\emph
  {\bibinfo {booktitle} {Quantum Probability and Related Topics}}},\
  Vol.~\bibinfo {volume} {27}\ (\bibinfo  {publisher} {World Scientific},\
  \bibinfo {year} {2011})\ pp.\ \bibinfo {pages} {261--281}\BibitemShut
  {NoStop}%
\bibitem [{\citenamefont {Paur}\ \emph {et~al.}(2016)\citenamefont {Paur},
  \citenamefont {Stoklasa}, \citenamefont {Hradil}, \citenamefont
  {Sanchez-Soto},\ and\ \citenamefont {Rehacek}}]{Paur:2016aa}%
  \BibitemOpen
  \bibfield  {author} {\bibinfo {author} {\bibfnamefont {M.}~\bibnamefont
  {Paur}}, \bibinfo {author} {\bibfnamefont {B.}~\bibnamefont {Stoklasa}},
  \bibinfo {author} {\bibfnamefont {Z.}~\bibnamefont {Hradil}}, \bibinfo
  {author} {\bibfnamefont {L.~L.}\ \bibnamefont {Sanchez-Soto}}, \ and\
  \bibinfo {author} {\bibfnamefont {J.}~\bibnamefont {Rehacek}},\ }\href
  {\doibase 10.1364/OPTICA.3.001144} {\bibfield  {journal} {\bibinfo  {journal}
  {Optica}\ }\textbf {\bibinfo {volume} {3}},\ \bibinfo {pages} {1144}
  (\bibinfo {year} {2016})}\BibitemShut {NoStop}%
\bibitem [{\citenamefont {Tham}\ \emph {et~al.}(2016)\citenamefont {Tham},
  \citenamefont {Ferretti},\ and\ \citenamefont {Steinberg}}]{Tham:2016aa}%
  \BibitemOpen
  \bibfield  {author} {\bibinfo {author} {\bibfnamefont {W.~K.}\ \bibnamefont
  {Tham}}, \bibinfo {author} {\bibfnamefont {H.}~\bibnamefont {Ferretti}}, \
  and\ \bibinfo {author} {\bibfnamefont {A.~M.}\ \bibnamefont {Steinberg}},\
  }\href {http://arxiv.org/pdf/1606.02666.pdf} {\bibfield  {journal} {\bibinfo
  {journal} {Phys. Rev. Lett.}\ }\textbf {\bibinfo {volume} {118}},\ \bibinfo
  {pages} {070801} (\bibinfo {year} {2016})}\BibitemShut {NoStop}%
\bibitem [{\citenamefont {Yang}\ \emph {et~al.}(2016)\citenamefont {Yang},
  \citenamefont {Taschilina}, \citenamefont {Moiseev}, \citenamefont {Simon},\
  and\ \citenamefont {Lvovsky}}]{Yang:2016aa}%
  \BibitemOpen
  \bibfield  {author} {\bibinfo {author} {\bibfnamefont {F.}~\bibnamefont
  {Yang}}, \bibinfo {author} {\bibfnamefont {A.}~\bibnamefont {Taschilina}},
  \bibinfo {author} {\bibfnamefont {E.~S.}\ \bibnamefont {Moiseev}}, \bibinfo
  {author} {\bibfnamefont {C.}~\bibnamefont {Simon}}, \ and\ \bibinfo {author}
  {\bibfnamefont {A.~I.}\ \bibnamefont {Lvovsky}},\ }\href {\doibase
  10.1364/OPTICA.3.001148} {\bibfield  {journal} {\bibinfo  {journal} {Optica}\
  }\textbf {\bibinfo {volume} {3}},\ \bibinfo {pages} {1148} (\bibinfo {year}
  {2016})}\BibitemShut {NoStop}%
\bibitem [{\citenamefont {Matsumoto}(2002)}]{Matsumoto:2002aa}%
  \BibitemOpen
  \bibfield  {author} {\bibinfo {author} {\bibfnamefont {K.}~\bibnamefont
  {Matsumoto}},\ }\href {http://stacks.iop.org/0305-4470/35/i=13/a=307}
  {\bibfield  {journal} {\bibinfo  {journal} {J. Phys. A: Math. Gen.}\ }\textbf
  {\bibinfo {volume} {35}},\ \bibinfo {pages} {3111} (\bibinfo {year}
  {2002})}\BibitemShut {NoStop}%
\bibitem [{\citenamefont {Pezz{\`e}}\ \emph {et~al.}(2017)\citenamefont
  {Pezz{\`e}}, \citenamefont {Ciampini}, \citenamefont {Spagnolo},
  \citenamefont {Humphreys}, \citenamefont {Datta}, \citenamefont {Walmsley},
  \citenamefont {Barbieri}, \citenamefont {Sciarrino},\ and\ \citenamefont
  {Smerzi}}]{Pezze:2017aa}%
  \BibitemOpen
  \bibfield  {author} {\bibinfo {author} {\bibfnamefont {L.}~\bibnamefont
  {Pezz{\`e}}}, \bibinfo {author} {\bibfnamefont {M.~A.}\ \bibnamefont
  {Ciampini}}, \bibinfo {author} {\bibfnamefont {N.}~\bibnamefont {Spagnolo}},
  \bibinfo {author} {\bibfnamefont {P.~C.}\ \bibnamefont {Humphreys}}, \bibinfo
  {author} {\bibfnamefont {A.}~\bibnamefont {Datta}}, \bibinfo {author}
  {\bibfnamefont {I.~A.}\ \bibnamefont {Walmsley}}, \bibinfo {author}
  {\bibfnamefont {M.}~\bibnamefont {Barbieri}}, \bibinfo {author}
  {\bibfnamefont {F.}~\bibnamefont {Sciarrino}}, \ and\ \bibinfo {author}
  {\bibfnamefont {A.}~\bibnamefont {Smerzi}},\ }\href {\doibase
  10.1103/PhysRevLett.119.130504} {\bibfield  {journal} {\bibinfo  {journal}
  {Phys. Rev. Lett.}\ }\textbf {\bibinfo {volume} {119}},\ \bibinfo {pages}
  {130504} (\bibinfo {year} {2017})}\BibitemShut {NoStop}%
\bibitem [{\citenamefont {Szczykulska}\ \emph {et~al.}(2016)\citenamefont
  {Szczykulska}, \citenamefont {Baumgratz},\ and\ \citenamefont
  {Datta}}]{Szczykulska:2016aa}%
  \BibitemOpen
  \bibfield  {author} {\bibinfo {author} {\bibfnamefont {M.}~\bibnamefont
  {Szczykulska}}, \bibinfo {author} {\bibfnamefont {T.}~\bibnamefont
  {Baumgratz}}, \ and\ \bibinfo {author} {\bibfnamefont {A.}~\bibnamefont
  {Datta}},\ }\href {\doibase 10.1080/23746149.2016.1230476} {\bibfield
  {journal} {\bibinfo  {journal} {Adv. Phys. X}\ }\textbf {\bibinfo {volume}
  {1}},\ \bibinfo {pages} {621} (\bibinfo {year} {2016})}\BibitemShut {NoStop}%
\bibitem [{\citenamefont {Albarelli}\ \emph {et~al.}(2020)\citenamefont
  {Albarelli}, \citenamefont {Barbieri}, \citenamefont {Genoni},\ and\
  \citenamefont {Gianani}}]{Albarelli:2020aa}%
  \BibitemOpen
  \bibfield  {author} {\bibinfo {author} {\bibfnamefont {F.}~\bibnamefont
  {Albarelli}}, \bibinfo {author} {\bibfnamefont {M.}~\bibnamefont {Barbieri}},
  \bibinfo {author} {\bibfnamefont {M.~G.}\ \bibnamefont {Genoni}}, \ and\
  \bibinfo {author} {\bibfnamefont {I.}~\bibnamefont {Gianani}},\ }\href
  {\doibase https://doi.org/10.1016/j.physleta.2020.126311} {\bibfield
  {journal} {\bibinfo  {journal} {Phys. Lett. A}\ }\textbf {\bibinfo {volume}
  {384}},\ \bibinfo {pages} {126311} (\bibinfo {year} {2020})}\BibitemShut
  {NoStop}%
\bibitem [{\citenamefont {Vidrighin}\ \emph {et~al.}(2014)\citenamefont
  {Vidrighin}, \citenamefont {Donati}, \citenamefont {Genoni}, \citenamefont
  {Jin}, \citenamefont {Kolthammer}, \citenamefont {Kim}, \citenamefont
  {Datta}, \citenamefont {Barbieri},\ and\ \citenamefont
  {Walmsley}}]{Vidrighin:2014aa}%
  \BibitemOpen
  \bibfield  {author} {\bibinfo {author} {\bibfnamefont {M.~D.}\ \bibnamefont
  {Vidrighin}}, \bibinfo {author} {\bibfnamefont {G.}~\bibnamefont {Donati}},
  \bibinfo {author} {\bibfnamefont {M.~G.}\ \bibnamefont {Genoni}}, \bibinfo
  {author} {\bibfnamefont {X.~M.}\ \bibnamefont {Jin}}, \bibinfo {author}
  {\bibfnamefont {W.~S.}\ \bibnamefont {Kolthammer}}, \bibinfo {author}
  {\bibfnamefont {M.~S.}\ \bibnamefont {Kim}}, \bibinfo {author} {\bibfnamefont
  {A.}~\bibnamefont {Datta}}, \bibinfo {author} {\bibfnamefont
  {M.}~\bibnamefont {Barbieri}}, \ and\ \bibinfo {author} {\bibfnamefont
  {I.~A.}\ \bibnamefont {Walmsley}},\ }\href {\doibase 10.1038/ncomms4532}
  {\bibfield  {journal} {\bibinfo  {journal} {Nat. Commun.}\ }\textbf {\bibinfo
  {volume} {5}},\ \bibinfo {pages} {3532} (\bibinfo {year} {2014})}\BibitemShut
  {NoStop}%
\bibitem [{\citenamefont {Polino}\ \emph {et~al.}(2019)\citenamefont {Polino},
  \citenamefont {Riva}, \citenamefont {Valeri}, \citenamefont {Silvestri},
  \citenamefont {Corrielli}, \citenamefont {Crespi}, \citenamefont {Spagnolo},
  \citenamefont {Osellame},\ and\ \citenamefont {Sciarrino}}]{Polino:2019aa}%
  \BibitemOpen
  \bibfield  {author} {\bibinfo {author} {\bibfnamefont {E.}~\bibnamefont
  {Polino}}, \bibinfo {author} {\bibfnamefont {M.}~\bibnamefont {Riva}},
  \bibinfo {author} {\bibfnamefont {M.}~\bibnamefont {Valeri}}, \bibinfo
  {author} {\bibfnamefont {R.}~\bibnamefont {Silvestri}}, \bibinfo {author}
  {\bibfnamefont {G.}~\bibnamefont {Corrielli}}, \bibinfo {author}
  {\bibfnamefont {A.}~\bibnamefont {Crespi}}, \bibinfo {author} {\bibfnamefont
  {N.}~\bibnamefont {Spagnolo}}, \bibinfo {author} {\bibfnamefont
  {R.}~\bibnamefont {Osellame}}, \ and\ \bibinfo {author} {\bibfnamefont
  {F.}~\bibnamefont {Sciarrino}},\ }\href {\doibase 10.1364/OPTICA.6.000288}
  {\bibfield  {journal} {\bibinfo  {journal} {Optica}\ }\textbf {\bibinfo
  {volume} {6}},\ \bibinfo {pages} {288} (\bibinfo {year} {2019})}\BibitemShut
  {NoStop}%
\bibitem [{\citenamefont {Watts}\ and\ \citenamefont
  {Rosenberg}(2017)}]{Watts:2017aa}%
  \BibitemOpen
  \bibfield  {author} {\bibinfo {author} {\bibfnamefont {S.}~\bibnamefont
  {Watts}}\ and\ \bibinfo {author} {\bibfnamefont {L.}~\bibnamefont
  {Rosenberg}},\ }in\ \href {\doibase 10.1049/cp.2017.0497} {\emph {\bibinfo
  {booktitle} {Proc. Int. Conf. Radar Syst.}}}\ (\bibinfo {year} {2017})\ pp.\
  \bibinfo {pages} {1--6}\BibitemShut {NoStop}%
\bibitem [{\citenamefont {Rehacek}\ \emph
  {et~al.}(2017{\natexlab{a}})\citenamefont {Rehacek}, \citenamefont
  {Pa{\'u}r}, \citenamefont {Stoklasa}, \citenamefont {Hradil},\ and\
  \citenamefont {S{\'a}nchez-Soto}}]{Rehacek:2017aa}%
  \BibitemOpen
  \bibfield  {author} {\bibinfo {author} {\bibfnamefont {J.}~\bibnamefont
  {Rehacek}}, \bibinfo {author} {\bibfnamefont {M.}~\bibnamefont {Pa{\'u}r}},
  \bibinfo {author} {\bibfnamefont {B.}~\bibnamefont {Stoklasa}}, \bibinfo
  {author} {\bibfnamefont {Z.}~\bibnamefont {Hradil}}, \ and\ \bibinfo {author}
  {\bibfnamefont {L.~L.}\ \bibnamefont {S{\'a}nchez-Soto}},\ }\href {\doibase
  10.1364/OL.42.000231} {\bibfield  {journal} {\bibinfo  {journal} {Opt.
  Lett.}\ }\textbf {\bibinfo {volume} {42}},\ \bibinfo {pages} {231} (\bibinfo
  {year} {2017}{\natexlab{a}})}\BibitemShut {NoStop}%
\bibitem [{\citenamefont {{\v R}eh{\'a}{\v c}ek}\ \emph
  {et~al.}(2018)\citenamefont {{\v R}eh{\'a}{\v c}ek}, \citenamefont {Hradil},
  \citenamefont {Koutn{\'y}}, \citenamefont {Grover}, \citenamefont {Krzic},\
  and\ \citenamefont {S{\'a}nchez-Soto}}]{Rehacek:2018aa}%
  \BibitemOpen
  \bibfield  {author} {\bibinfo {author} {\bibfnamefont {J.}~\bibnamefont {{\v
  R}eh{\'a}{\v c}ek}}, \bibinfo {author} {\bibfnamefont {Z.}~\bibnamefont
  {Hradil}}, \bibinfo {author} {\bibfnamefont {D.}~\bibnamefont {Koutn{\'y}}},
  \bibinfo {author} {\bibfnamefont {J.}~\bibnamefont {Grover}}, \bibinfo
  {author} {\bibfnamefont {A.}~\bibnamefont {Krzic}}, \ and\ \bibinfo {author}
  {\bibfnamefont {L.~L.}\ \bibnamefont {S{\'a}nchez-Soto}},\ }\href {\doibase
  10.1103/PhysRevA.98.012103} {\bibfield  {journal} {\bibinfo  {journal} {Phys.
  Rev. A}\ }\textbf {\bibinfo {volume} {98}},\ \bibinfo {pages} {012103}
  (\bibinfo {year} {2018})}\BibitemShut {NoStop}%
\bibitem [{\citenamefont {Eckstein}\ \emph {et~al.}(2011)\citenamefont
  {Eckstein}, \citenamefont {Brecht},\ and\ \citenamefont
  {Silberhorn}}]{Eckstein:2011aa}%
  \BibitemOpen
  \bibfield  {author} {\bibinfo {author} {\bibfnamefont {A.}~\bibnamefont
  {Eckstein}}, \bibinfo {author} {\bibfnamefont {B.}~\bibnamefont {Brecht}}, \
  and\ \bibinfo {author} {\bibfnamefont {C.}~\bibnamefont {Silberhorn}},\
  }\href@noop {} {\bibfield  {journal} {\bibinfo  {journal} {Opt. Express}\
  }\textbf {\bibinfo {volume} {19}},\ \bibinfo {pages} {13770} (\bibinfo {year}
  {2011})}\BibitemShut {NoStop}%
\bibitem [{\citenamefont {Brecht}\ \emph {et~al.}(2014)\citenamefont {Brecht},
  \citenamefont {Eckstein}, \citenamefont {Ricken}, \citenamefont {Quiring},
  \citenamefont {Suche}, \citenamefont {Sansoni},\ and\ \citenamefont
  {Silberhorn}}]{brecht2014demonstration}%
  \BibitemOpen
  \bibfield  {author} {\bibinfo {author} {\bibfnamefont {B.}~\bibnamefont
  {Brecht}}, \bibinfo {author} {\bibfnamefont {A.}~\bibnamefont {Eckstein}},
  \bibinfo {author} {\bibfnamefont {R.}~\bibnamefont {Ricken}}, \bibinfo
  {author} {\bibfnamefont {V.}~\bibnamefont {Quiring}}, \bibinfo {author}
  {\bibfnamefont {H.}~\bibnamefont {Suche}}, \bibinfo {author} {\bibfnamefont
  {L.}~\bibnamefont {Sansoni}}, \ and\ \bibinfo {author} {\bibfnamefont
  {C.}~\bibnamefont {Silberhorn}},\ }\href@noop {} {\bibfield  {journal}
  {\bibinfo  {journal} {Physical Review A}\ }\textbf {\bibinfo {volume} {90}},\
  \bibinfo {pages} {030302} (\bibinfo {year} {2014})}\BibitemShut {NoStop}%
\bibitem [{\citenamefont {Ansari}\ \emph
  {et~al.}(2018{\natexlab{a}})\citenamefont {Ansari}, \citenamefont {Donohue},
  \citenamefont {Allgaier}, \citenamefont {Sansoni}, \citenamefont {Brecht},
  \citenamefont {Roslund}, \citenamefont {Treps}, \citenamefont {Harder},\ and\
  \citenamefont {Silberhorn}}]{ansari2018tomography}%
  \BibitemOpen
  \bibfield  {author} {\bibinfo {author} {\bibfnamefont {V.}~\bibnamefont
  {Ansari}}, \bibinfo {author} {\bibfnamefont {J.~M.}\ \bibnamefont {Donohue}},
  \bibinfo {author} {\bibfnamefont {M.}~\bibnamefont {Allgaier}}, \bibinfo
  {author} {\bibfnamefont {L.}~\bibnamefont {Sansoni}}, \bibinfo {author}
  {\bibfnamefont {B.}~\bibnamefont {Brecht}}, \bibinfo {author} {\bibfnamefont
  {J.}~\bibnamefont {Roslund}}, \bibinfo {author} {\bibfnamefont
  {N.}~\bibnamefont {Treps}}, \bibinfo {author} {\bibfnamefont
  {G.}~\bibnamefont {Harder}}, \ and\ \bibinfo {author} {\bibfnamefont
  {C.}~\bibnamefont {Silberhorn}},\ }\href@noop {} {\bibfield  {journal}
  {\bibinfo  {journal} {Physical review letters}\ }\textbf {\bibinfo {volume}
  {120}},\ \bibinfo {pages} {213601} (\bibinfo {year}
  {2018}{\natexlab{a}})}\BibitemShut {NoStop}%
\bibitem [{\citenamefont {Silver}\ \emph {et~al.}(2019)\citenamefont {Silver},
  \citenamefont {Huang}, \citenamefont {Langrock}, \citenamefont {Fejer},
  \citenamefont {Kumar},\ and\ \citenamefont {Kanter}}]{Silver:2019aa}%
  \BibitemOpen
  \bibfield  {author} {\bibinfo {author} {\bibfnamefont {M.}~\bibnamefont
  {Silver}}, \bibinfo {author} {\bibfnamefont {Y.}~\bibnamefont {Huang}},
  \bibinfo {author} {\bibfnamefont {C.}~\bibnamefont {Langrock}}, \bibinfo
  {author} {\bibfnamefont {M.~M.}\ \bibnamefont {Fejer}}, \bibinfo {author}
  {\bibfnamefont {P.}~\bibnamefont {Kumar}}, \ and\ \bibinfo {author}
  {\bibfnamefont {G.~S.}\ \bibnamefont {Kanter}},\ }\bibfield  {booktitle}
  {\emph {\bibinfo {booktitle} {IEEE Photonics Technology Letters}},\ }\href
  {\doibase 10.1109/LPT.2019.2944636} {\bibfield  {journal} {\bibinfo
  {journal} {IEEE Photonics Tech. Lett.}\ }\textbf {\bibinfo {volume} {31}},\
  \bibinfo {pages} {1749} (\bibinfo {year} {2019})}\BibitemShut {NoStop}%
\bibitem [{\citenamefont {Rehacek}\ \emph
  {et~al.}(2017{\natexlab{b}})\citenamefont {Rehacek}, \citenamefont {Hradil},
  \citenamefont {Stoklasa}, \citenamefont {Pa{\'u}r}, \citenamefont {Grover},
  \citenamefont {Krzic},\ and\ \citenamefont
  {S{\'a}nchez-Soto}}]{Rehacek:2017ab}%
  \BibitemOpen
  \bibfield  {author} {\bibinfo {author} {\bibfnamefont {J.}~\bibnamefont
  {Rehacek}}, \bibinfo {author} {\bibfnamefont {Z.}~\bibnamefont {Hradil}},
  \bibinfo {author} {\bibfnamefont {B.}~\bibnamefont {Stoklasa}}, \bibinfo
  {author} {\bibfnamefont {M.}~\bibnamefont {Pa{\'u}r}}, \bibinfo {author}
  {\bibfnamefont {J.}~\bibnamefont {Grover}}, \bibinfo {author} {\bibfnamefont
  {A.}~\bibnamefont {Krzic}}, \ and\ \bibinfo {author} {\bibfnamefont {L.~L.}\
  \bibnamefont {S{\'a}nchez-Soto}},\ }\href {\doibase
  10.1103/PhysRevA.96.062107} {\bibfield  {journal} {\bibinfo  {journal} {Phys.
  Rev. A}\ }\textbf {\bibinfo {volume} {96}},\ \bibinfo {pages} {062107}
  (\bibinfo {year} {2017}{\natexlab{b}})}\BibitemShut {NoStop}%
\bibitem [{\citenamefont {Eldar}(2006)}]{Eldar:2006aa}%
  \BibitemOpen
  \bibfield  {author} {\bibinfo {author} {\bibfnamefont {Y.~C.}\ \bibnamefont
  {Eldar}},\ }\href {\doibase 10.1109/TSP.2006.877648} {\bibfield  {journal}
  {\bibinfo  {journal} {IEEE Trans. Signal Process.}\ }\textbf {\bibinfo
  {volume} {54}},\ \bibinfo {pages} {2943} (\bibinfo {year}
  {2006})}\BibitemShut {NoStop}%
\bibitem [{\citenamefont {Ansari}\ \emph
  {et~al.}(2018{\natexlab{b}})\citenamefont {Ansari}, \citenamefont {Donohue},
  \citenamefont {Brecht},\ and\ \citenamefont
  {Silberhorn}}]{ansari2018tailoring}%
  \BibitemOpen
  \bibfield  {author} {\bibinfo {author} {\bibfnamefont {V.}~\bibnamefont
  {Ansari}}, \bibinfo {author} {\bibfnamefont {J.~M.}\ \bibnamefont {Donohue}},
  \bibinfo {author} {\bibfnamefont {B.}~\bibnamefont {Brecht}}, \ and\ \bibinfo
  {author} {\bibfnamefont {C.}~\bibnamefont {Silberhorn}},\ }\href@noop {}
  {\bibfield  {journal} {\bibinfo  {journal} {Optica}\ }\textbf {\bibinfo
  {volume} {5}},\ \bibinfo {pages} {534} (\bibinfo {year}
  {2018}{\natexlab{b}})}\BibitemShut {NoStop}%
\end{thebibliography}

%

\end{document}